\documentclass[twocolumn]{aa}
\usepackage{psfig,subfigure,graphicx,natbib}

\begin{document}

\title{Radio emission from exoplanets: the role of the stellar coronal density and magnetic field strength}

\author
{ M. Jardine \inst{1} \and A. Collier Cameron \inst{1} }

\offprints{M. Jardine}

\institute{
SUPA, School of Physics and Astronomy, Univ.\ of St~Andrews, 
St~Andrews, 
Scotland KY16 9SS \\
email{mmj@st-andrews.ac.uk}
}

\date{Received; accepted 2008}

\abstract
{The search for radio emission from extra-solar planets has so far been unsuccessful. Much of the effort in modelling the predicted emission has been based on the analogy with the well-known emission from Jupiter. Unlike Jupiter, however, many of the targets of these radio searches are so close to their parent stars that they may well lie inside the stellar magnetosphere.}
{For these close-in planets we determine which physical processes dominate the radio emission and compare our results to those for large-orbit planets that are immersed in the stellar wind.}
{We have modelled the reconnection of the stellar and planetary magnetic fields. We calculate the extent of the planetary magnetosphere if it is in pressure balance with its surroundings and determine the conditions under which reconnection of the stellar and planetary magnetic fields could provide the accelerated electrons necessary for the predicted radio emission. }
{We show that received radio fluxes of tens of mJy are possible for exoplanets in the solar neighbourhood that are close to their parent stars if their stars have surface field strengths above 1-10G. We show that for these close-in planets,  the power of the radio emission depends principally on the ratio $(N_c/B_\star^{1/3})^2$ where $N_c$ is the density at the base of the stellar corona, and $B_\star$ is the stellar surface magnetic field strength.}
{Radio emission is most likely to be detected from planets around stars with high-density coronae, which are therefore likely to be bright X-ray sources. The dependence of stellar coronal density on stellar rotation rate and effective temperature is crucial in predicting radio fluxes from exoplanets.}


\maketitle

%--------------------------------------------------------------------
\section{Introduction}

Searches for the radio signatures of extra-solar planets have not yet yielded any detections. The motivation for these searches was the well-known decametre emission from Jupiter which can outshine the radio emission from the quiet Sun at these wavelengths by some four orders of magnitude \citep{farrell_radio_planets_99}. This emission is believed to originate from non-thermal electrons travelling down the converging field lines at the planet's magnetic poles  \citep{dulk85,zarka_98}. There are several possible sources for these electrons, including currents generated by corotation breakdown inside the middle magnetosphere, the Io-Jupiter interaction and the reconnection of Jupiter's magnetic field with the solar wind magnetic field \citep{cowley_01,saur_04}. Here we focus on the third possibility since it is common to all the magnetised planets. If these electrons are accelerated to energies greater than 1-10 keV, they  can be unstable to the electron cyclotron maser instability and as a result they emit at the local electron gyrofrequency $f \mbox{[MHz] = 2.8B(G)}$. The observed peak of Jupiter's emission at 39.5MHz suggests a field strength of 14.5G at the site of the emission close to Jupiter's surface  \citep{connerney_98}. The proposal that such a situation might occur in some of the known ``Hot Jupiters" led to several searches but none of these has revealed the expected emission  \citep{ryabov_04}. A comprehensive review of the current status of theory and observation is given in \citet{zarka_07}. 

One of the most promising reasons for this lack of detections is that tides raised on planets orbiting close to their host stars rapidly synchronise the axial spin with the orbital motion. In this case, the planet may be rotating slowly, producing only a weak magnetic field.  Since the emission frequency is proportional to the magnetic field strength, this could result in emission at unobservably low frequencies. The other synchronisation that can take place is between the spin of the star and the orbit of the planet. This timescale is set by torques due to the tidal bulges raised on the star by the planet (see, e.g. \citet{zahn77}) and is generally expected to be longer than the star's main-sequence lifetime even for close-orbiting gas-giant planets.  The observed spin rates of planet-host stars are consistent with this expectation. A notable exception is the late F star tau Boo, which rotates synchronously with the 3.3-day orbit of its planet, whose mass is sufficient to lock the star's rotation within its main-sequence lifetime  \citep{marcy_51peg_97,lubow_spindown_97}. More recently \citet{dobbs_dixon_planets_04} have considered the combined effects of tidal torques and angular momentum loss from the star in the form of a wind. They  suggest that synchronisation is more likely among F stars than among the lower mass G and K stars.

The nature of the stellar wind may also have an impact on the expected radio emission, since the wind may compress the stellar magnetosphere and also may strip particles out of the planetary exosphere \citep{griessmeier_planet_tides_04,griessmeier_planet_winds_05,jaritz_planet_roche_05,lipatov_3Dhybrid_05,stevens_planet_winds_05}.  Fluctuations in the stellar wind pressure and speed that occur as a result of the stellar equivalent of solar coronal mass ejections may, however, enhance any planetary radio emission \citep{khodachenko_CME_07}.  A recent search for radio emission from systems thought to be embedded in strong stellar winds has yielded no detections, but tight upper limits \citep{stevens_planet_winds_05,george_radio_planets_07}.

It is clear from this modeling that conditions both exterior to the planet (such as the speed, density and magnetic energy of the stellar wind) and within the planet itself (such as the nature of the dynamo-generated magnetic field) may have an impact on the detectability of exoplanetary radio emission. There is another factor, however, which is the process by which the power carried in the stellar wind is converted into radio emission. Within our own solar system, there is a clear scaling between the magnetic or kinetic power in the solar wind incident on a planet and the observed output radio power \citep{farrell_radio_planets_99,zarka_01,zarka_07}. This  ``radio-magnetic Bode's law'' can be used to predict the radio emission from a variety of stellar wind or planetary magnetosphere conditions. 

In this paper we take a complementary approach. Rather than focussing on the physics of the stellar wind or the planet, we choose simple models for these and focus instead on the process of energy conversion. Our aim is to examine how the radio power scales with different stellar parameters, and  to determine if this scaling is the same for all exoplanets.

%-------------------  Fig 1 -----------------------------------------
 \begin{figure}
  \begin{center}
   \psfig{figure=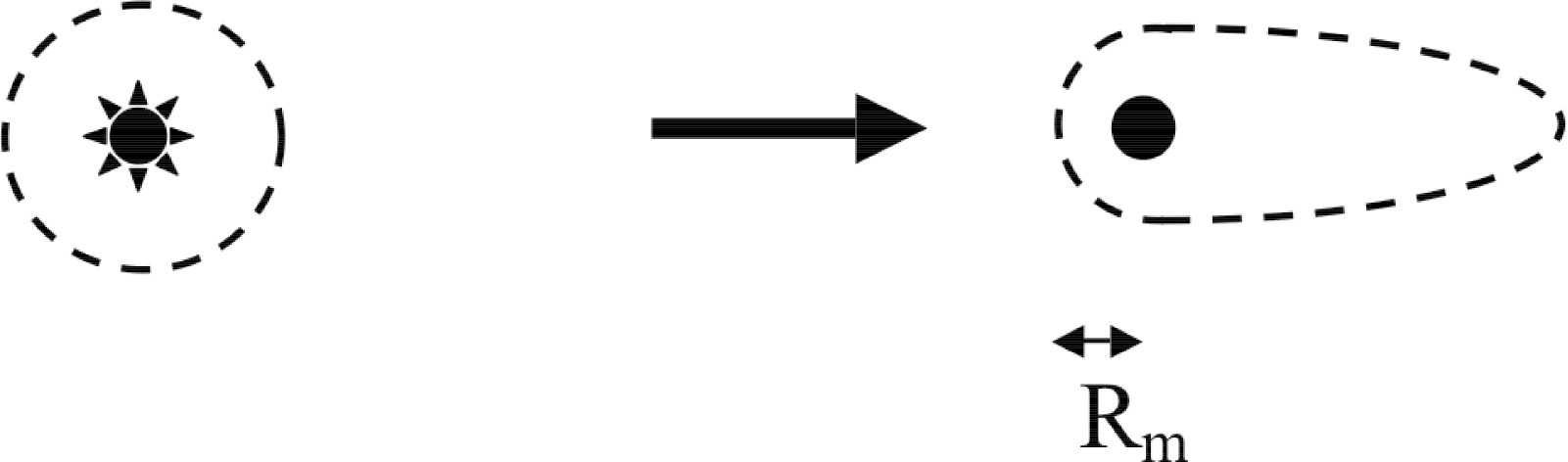,width=2.5in}
   \caption{When the planet is far from the star, the stellar wind elongates the planetary magnetosphere in the direction away from the star. $R_{\rm m}$ is the extent of the planetary magnetosphere which is determine by the balance of pressures between the planetary and stellar plasmas. The shapes of the stellar and planetary magnetosphere are shown as dashed lines (not to scale).}
   \label{cartoon_1}
  \end{center}
\end{figure}
%------------------------------------------------------------
%---------------------Fig 2------------------------------
 \begin{figure}
  \begin{center}
     \psfig{figure=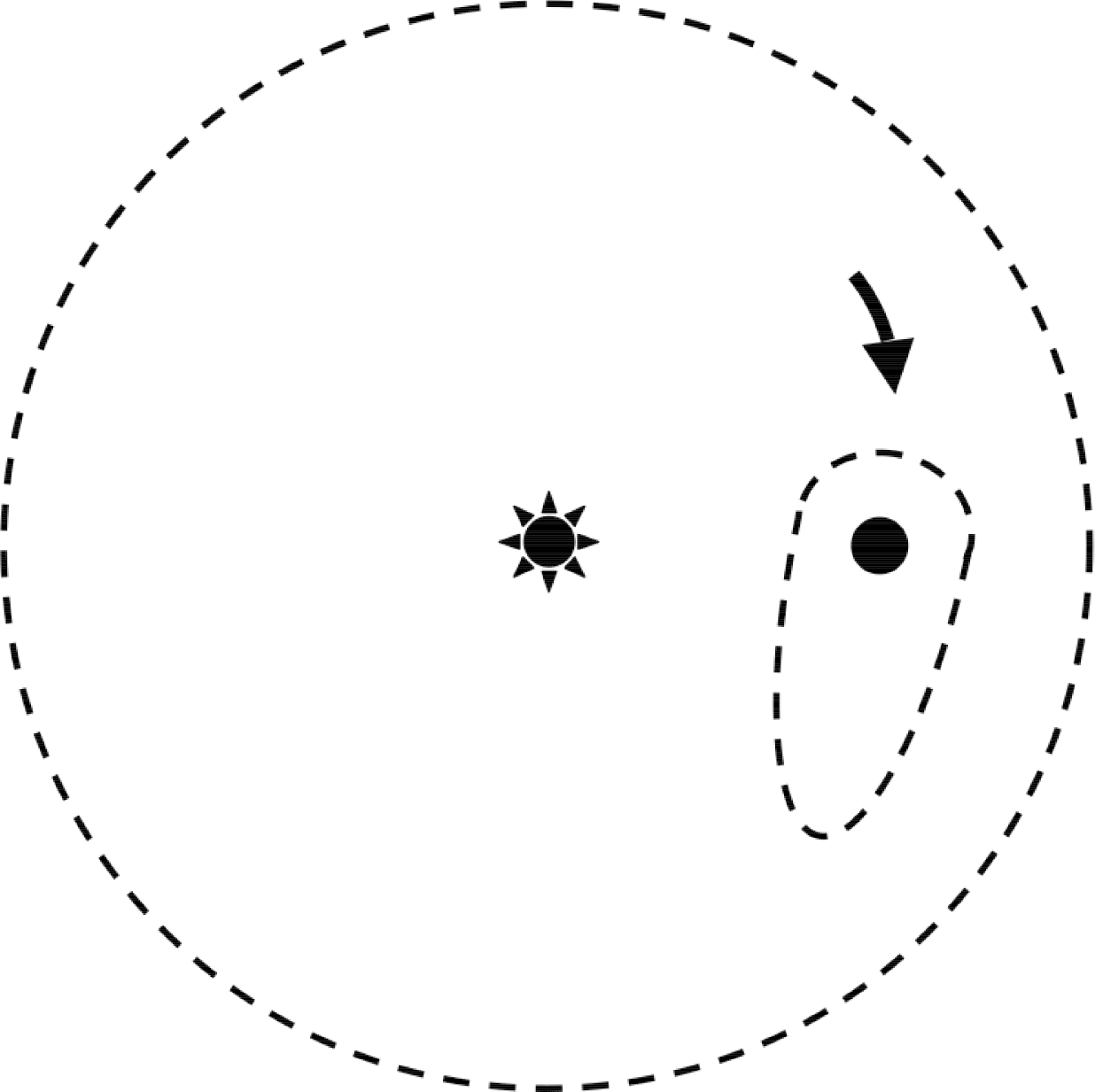,width=2.in}
   \caption{When the planet is within the stellar magnetosphere, the effective headwind due to the relative motion between the planet and the stellar magnetosphere  elongates the planetary magnetosphere in the azimuthal direction (see also \citet{zarka_01}).}
   \label{cartoon_2}
  \end{center}
\end{figure}
%------------------------------------------------------------
\section{The model}
We begin by considering the interaction between the planet and its immediate environment. In the case of the outer planets, this takes the form of a stellar wind impacting the planetary magnetosphere. This wind carries both magnetic and kinetic energy, but it is the magnetic component that we are particularly interested in. Following previous work (e.g. \citet{zarka_01,zarka_07}),  we can calculate the rate at which the stellar wind delivers magnetic energy to the planetary magnetosphere by calculating the Poynting flux $(\bf{E} \times {\bf B}/ \mu)$ that is incident on the cross-sectional area $A$ of the planetary magnetosphere. We can approximate this maximum available power as $P = AvB^2/\mu$. This depends on the cross-section of the planetary magnetosphere $A = \pi R_m^2$ the wind velocity $v$ and the stellar field strength $B$ as measured at the planetary magnetosphere. This maximum available power sets an upper limit on the power that could be detected in radio emission.

The situation for many ``Hot Jupiters" is, however, somewhat different. For these planets, the orbital radius is extremely small - so much so that the planet may in fact be inside the stellar magnetosphere, where the stellar magnetic field lines are closed and confine the coronal gas. The maximum extent of the stellar magnetosphere is expected to increase with stellar activity (and hence with stellar rotation rate). Thus from a value of 2.5R$_\odot$ seen in eclipse images of the Sun \citep{altschuler69,badalian_Kcorona_86}, it may increase to $\approx$19R$_\odot$ or 0.09AU for a star with rotation period about 6 days that is 10 times more active than the Sun in terms of surface magnetic flux density or soft X-ray luminosity \citep{schrijver01,schrijver_asterosphere_03}. Given the youth of some exoplanetary systems, many planet host stars may have magnetic activity strong enough to support magnetospheres extending to 0.1 AU or so. Hot Jupiters with orbital distance of order 0.05 AU will thus orbit within their host stars' magnetospheres. In this case, the velocity with which new magnetic field lines are carried into the interaction region is the relative velocity between the stellar and planetary magnetospheres.  In the frame of the planet, however, this would appear to be similar to the effect of a stellar wind.

The stellar wind therefore has two main effects on the planetary magnetosphere. The wind ram pressure compresses the planetary magnetosphere and elongates it as shown in Figs (\ref{cartoon_1}) and (\ref{cartoon_2}). The magnetic field that is carried in the wind is also pushed against the planetary magnetic field forming a narrow interaction region or {\em current sheet}. The magnetic field lines will pile up in this region until reconnection allows the field to dissipate. This process is observed both on the dayside and nightside of the Earth's magnetosphere and is related to the occurrence of aurorae as accelerated electrons stream along the Earth's magnetic field lines. 

The interaction of the planetary and stellar magnetic fields also generates an electric field at the interaction (or reconnection) site.  If this electric field has a component along the magnetic field lines, it will accelerate electrons along the magnetic field lines. These electrons will mostly be slowed down by collisions, but some fraction may escape and be freely accelerated. It is this small fraction of the electron population that may provide the pool of energetic electrons needed to supply the electron cyclotron maser instability that would result in observable radio emission. 

In those rare cases, such as the anomalously massive $\tau$ Boo b, where the stellar rotation  is synchronised with the orbital motion, then there will be no relative motion between the stellar field and the planetary field and so although reconnection may happen sporadically due to  the small velocity differences due to stellar differential rotation or  resistive instabilities  such as perhaps the tearing mode \citep{Furth63,cuntz_exoplanets_00},  it is unlikely to happen in the quasi-continuous manner that would be needed to power an observable radio signal. In the majority of observed cases, however, the star's rotation is not synchronous. New magnetic flux will then be brought into the interaction region at a rate determined by the velocity difference between the stellar and planetary fields. In this case the geometry of the interaction is altered and the elongation of the planetary magnetosphere will not be directed away from the star, but will trail the orbital motion ( see Figs. \ref{cartoon_1} and \ref{cartoon_2}).  

%--------------------Fig 3-------------------------
 \begin{figure}
  \begin{center}
       \psfig{figure=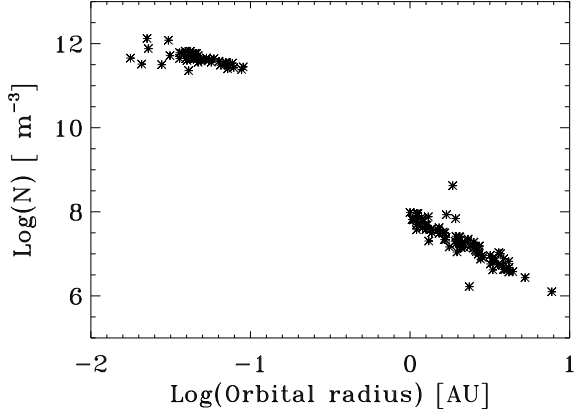,width=3.5in}
   \caption{Density of the gas in the region exterior to the magnetospheres of known exoplanets using orbital radii and stellar masses taken from The Extrasolar Planets Encyclopaedia at http://exoplanet.eu/index.php and assuming $R_\star \propto M_\star$. Close to the star, the planet is immersed in the the stellar magnetosphere, while far from the star, it is immersed in the stellar wind. At intermediate radii, planets may be in either regime and so no cases are plotted.}
   \label{density}
  \end{center}
\end{figure}
%------------------------------------------------------------
%--------------------Fig 4-------------------------------
 \begin{figure}
  \begin{center}
       \psfig{figure=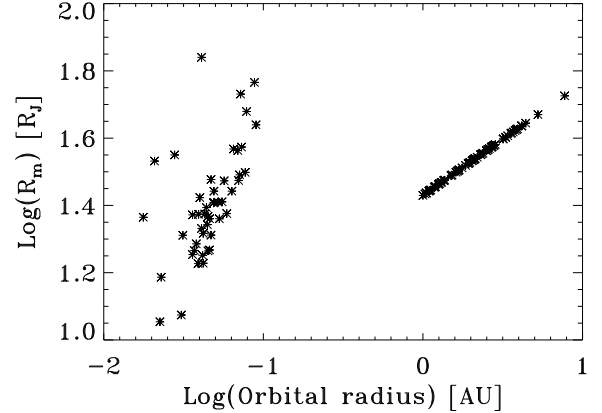,width=3.5in}
   \caption{Magnetospheric radius (in Jupiter radii) for known exoplanets as a function of orbital separation. A stellar field $B_\star =1$G is assumed.}
   \label{size}
  \end{center}
\end{figure}
%------------------------------------------------------------
\section{Magnetospheric radius}
%------------------------------------------------------------

The first step in calculating the radio power emitted by exoplanets is to calculate the cross-section of the planetary magnetosphere that intercepts the stellar wind. The magnetospheric radius is determined by pressure balance between the external medium and the planetary magnetosphere:
\begin{equation}
{B_p^2} =  {B_e^2} +\mu \rho_e v^2.
\label{press_bal}
\end{equation}
here $B_p$ is the magnetic field of the planet which we take to be 14.5G. We assume that this is a dipole, and so for a planet of radius $R_p$ this scales with distance as $(R/R_p)^{-3}$. The field in the external region $B_e$ is the stellar field. Close to the star, where the field lines are still closed, we assume that this field is also a dipole and so for a star of radius $R_\star$ this field scales with distance as $(R/R_\star)^{-3}$. At distance greater than that of the stellar magnetosphere, however, where the field lines have been opened up by the stellar wind, the field scales as $(R/R_\star)^{-2}$. For the inner planets, the velocity in (\ref{press_bal}) is the relative velocity between the star and planet:
\begin{eqnarray*}
v  & = & v_p - v_\star \\
             & = & 3 \times 10^4   \left[ \frac{R_{orb}}{0.03AU} \right]
                                                \left[ \frac{11d}{p_\star} \right]
                                                 \left[ \frac{p_\star}{p_{orb}}  - 1 \right] [\mbox{ms}^{-1}]
%\label{delta_v}  
\end{eqnarray*}
where we take the stellar rotation period to be 11 days (appropriate for HD189733 \citep{bouchy_hd189733_05}).
For the outer planets, this velocity is the wind velocity. This is an extremely difficult parameter to determine observationally, though use of hydrogen-wall absorption to deduce the wind ram pressure promises to be a useful tool \citep{wood_asterospheres_04}. Models of these winds also give very conflicting results \citep{stevens_planet_winds_05,holzwarth_coolwinds_07} and so, for the moment, we simply assume that the winds have reached their terminal velocity and so the wind density falls off with distance as $(R/R_\star)^{-2}$. We scale the stellar wind velocity to solar values, assuming a nominal solar wind speed  of 500kms$^{-1}$ and a density at 1AU of 1.7$\times 10^{-20}$kgm$^{-3}$ \citep{cox91}. This gives a number density 
\begin{eqnarray}
N & = & N_w(1AU)\left[ \frac{R_\star}{R_{orb}}\right]^2 \\
    & = &10^7 \left[\frac{R_\star}{R_\odot} \right]^2
                   \left[\frac{R_{orb}}{1AU}   \right] ^{-2}[\mbox{m}^{-3}].
\label{density_m}
\end{eqnarray}

For close-orbiting planets, however, the external density is not that of a wind, but that of the stellar magnetosphere. This is likely to be a function not only of the orbital radius but also the magnetospheric temperature and the stellar rotation rate (which influences the level of magnetic activity).  We assume that the stellar magnetosphere is in isothermal, hydrostatic balance and so obtain (see Fig. \ref{density})
\begin{equation}
N = N_0 \rm{exp} \left(  
                          -\frac{\mu m_H}{kT} \frac{GM_\star}{R_\star} 
                           \left[ 1-\frac{R_\star}{R_{orb}}
                           \right]
                              \right)[\mbox{m}^{-3}]
\label{density_w}
\end{equation}
with the number density at the base of the corona $N_0 = 4 \times 10^{14}$m$^{-3}$, the  temperature T=$1.4\times 10^6$K and the mean particle mass $\mu=0.6$ \citep{badalian_Kcorona_86}. We note that this gives a different scaling of the density $N$ with orbital radius in the two cases; for the outer planets, $N \propto R^{-2}$, whereas for the inner planets, $N\propto R^{-1}$.

Using these figures, we find, in common with previous studies \citep{cuntz_exoplanets_00, zarka_01,zarka_07} that for the more distant planets, the contribution of the external field to pressure balance is negligible. For the inner planets the opposite is true, and we may neglect the ram pressure associated with the relative motion of the planet and the star. This then gives, for the inner planets:
\begin{equation}
 \frac{R_m}{R_p} = 
      15.7  \left[ \frac{R_{orb}}{0.03AU} \right] \left[ \frac{B_p}{14.5G} \right]^{1/3} 
                    \left[ \frac{R_\star}{R_\odot} \right] ^{-1}\left[ \frac{B_\star}{1G} \right]^{-1/3},
\label{Rm_inner}
\end{equation}
and for the outer planets:
\begin{equation}
 \frac{R_m}{R_p} = 
      46 \left[ \frac{R_{orb}}{5AU} \right]^{1/3} \left[ \frac{B_p}{14.5G} \right]^{1/3} 
                    \left[ \frac{v_e}{500kms^{-1}} \right]^{-1/3}.
\label{Rm_outer}
\end{equation}
We show in Fig. (\ref{size}) the corresponding sizes of the planetary magnetospheres (assuming that all planets have a magnetic field similar to Jupiter's). While for planets far from their parent stars the magnetospheric size is independent of the the stellar field strength (being determined mainly by the wind ram pressure), for the close-in planets, the pressure of the stellar magnetic field can have a significant effect in compressing the planetary magnetosphere. If the stellar field strength is too high, the planetary magnetosphere may be totally crushed, but for intermediate stellar field strengths or stronger planetary fields the planet may still retain a magnetosphere even in a very close orbit.
%--------------------Fig 5----------------------
 \begin{figure}
  \begin{center}
       \psfig{figure=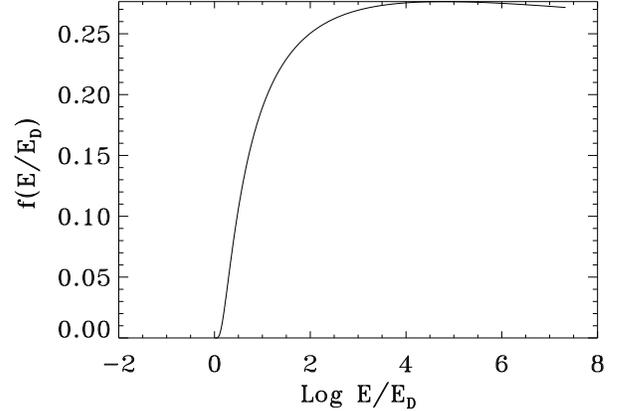,width=3.5in}
   \caption{The functional form of the dependence of the number of runaway electrons on the ratio of the applied electric field to the Dreicer field.}
   \label{f(ER)}
  \end{center}
\end{figure}
%------------------------------------------------------------
%----------------------Fig 6------------------------
 \begin{figure}
  \begin{center}
       \psfig{figure=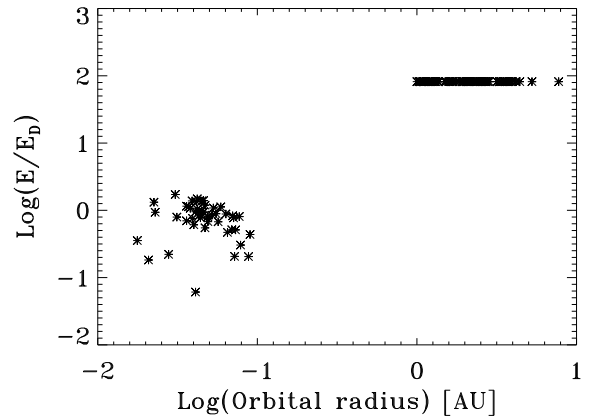,width=3.5in}
    \caption{Ratio of the electric field $E$ developed at the reconnection site to the Dreicer field $E_D$. Electron runaway occurs when $E>E_D$. A stellar field $B_\star =1$G is assumed.}
   \label{E_ED}
  \end{center}
\end{figure}
%------------------------------------------------------------

\section{The magnetic interaction between the star and the planet}

The interaction of the planetary magnetosphere with the external medium not only determines the size of the planetary magnetosphere but it also generates an electric field at the site where the magnetic fields of the planet and the star meet given by 
\begin{equation}
{\bf E} = -{\bf v}\times{\bf B} + {\bf j}/H
\end{equation}
where the current density $\bf{j} = \bf{\nabla}\times\bf{B}/\mu$.
Here we have scaled the electric field to its value far from the interaction site $E_0=-v_0B_0$ where $v_0$ and $B_0$ are the velocity and magnetic field strength. The current density is scaled to $B_0/\mu L$ where $L$ is the typical lengthscale for the interaction which we take to be 0.1$R_J$. The results are in fact fairly insensitive to the exact value of $L$ chosen. The magnetic Reynolds number $H=v_0 L/\eta$ where the Spitzer diffusivity is given by $\eta = 10^9 T^{-3/2}$m$^2$s$^{-1}$ \citep{priest84}. 

Thus, well away from the interaction site, the electric field is simply determined by the velocity and magnetic field strength and we have $\bf{E} = -\bf{v}\times\bf{B}$, while at the centre of the interaction region where $v=0$, the electric field is determined by the magnetic Reynolds number and we have ${\bf E}={\bf j}/H$.

It can be shown (see Appendix for details) that there is a  component of this electric field that is parallel to the magnetic field and so is capable of accelerating electrons along the magnetic field lines. As shown in the Appendix, this electric field ($E_{||}$) has a maximum at the interaction site between the fields of the star and the planet. Its magnitude falls away with increasing distance $z$ from this site as
\begin{equation}
{\bf E}_{||} = {\bf E}\cdot{\bf B} 
                        = {E}_{||}(0) e^{-z^2H}.
\end{equation}
It is the magnitude of this parallel electric field that is important in accelerating electrons in the interaction region. Its value is given simply by
\begin{equation}
    {E}_{||}(0) = -\frac{ 2\sqrt{2\pi}}
                             {\Gamma(1/4)} H^{-1/4} v_0 B_0.
\label{E_parallel}
\end{equation}
Hence, although there is a weak dependence on the magnetic Reynolds number (and hence on the temperature of the local plasma), the magnitude of this accelerating electric field is determined primarily by the magnetic field strength and the velocity at some distance from the interaction site.

\section{Power in accelerated electrons}
%-----------------Fig 7---------------------------
 \begin{figure}
  \begin{center}
       \psfig{figure=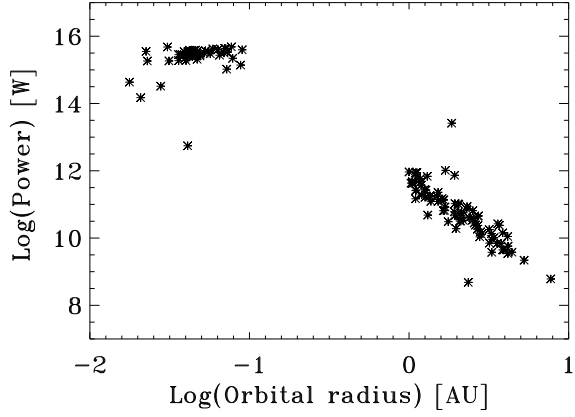,width=3.5in}
   \caption{Power in electrons accelerated by the interaction of the stellar and planetary magnetic fields. A stellar field $B_\star =1$G is assumed.}
   \label{power_all}
  \end{center}
\end{figure}
%------------------------------------------------------------
%-------------------Fig 8------------------------------
 \begin{figure}
  \begin{center}
       \psfig{figure=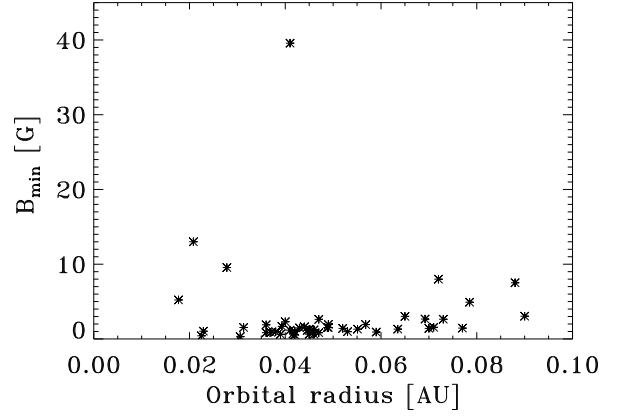,width=3.5in}
   \caption{The minimum stellar magnetic field needed to accelerate electrons to above 10keV in the interaction between the magnetic fields of the star and planet.}
   \label{B_min}
  \end{center}
\end{figure}
%------------------------------------------------------------
Once the electrons are accelerated by this field however, they may be slowed down by collisions with a collision frequency
\begin{equation}
\nu_c = 7.5\times 10^{-5}  T^{-3/2} n
 \end{equation}
 that depends on the local temperature $T$ and density $n$, here given in SI units \citep{kruskal64,tandberghanssen88}. However, if the electric field is stronger than a critical value known as the {\em Dreicer} field $E_D$, electrons may be freely accelerated and so will ``run away'' out of the thermal distribution. The Dreicer field is also a function of the local temperature and density and is given by
\begin{equation}
E_D = 18\times 10^{-12} n T^{-1}  [{\rm V m}^{-1}].
\label{dreicer}
\end{equation}
The number density of runaway electrons is then given by 
\begin{eqnarray}
N_{\rm run} & = & 0.35 n \nu_c f(E/E_D)  [{\rm m}^{-3}] \\
                    & = & 2.6 \times 10^{-5} n^2T^{-3/2}f(E/E_D)  [{\rm m}^{-3}] 
\label{N_acc}
\end{eqnarray}
where the function $f(E/E_D)$, shown in Fig. \ref{f(ER)}, is given by
\begin{equation}
f\left(\frac{E}{E_D}\right) = \left(\frac{E_D}{E}\right)^{3/8} {\rm exp} \left[-\left(\frac{2E_D}{E}\right)^{1/2} - \frac{E_D}{4E}\right].
\label{runaway}
\end{equation}
The behaviour of this function governs the way in which the electric field generated at the interaction site can affect the number density of electrons that run away and hence can power the radio emission. Once this imposed electric field is greater than the Dreicer field, there is a steep rise in $f(E/E_D)$ with increasing $E$. Once the ratio $E/E_D$ is sufficiently large, however, $f$ becomes insensitive to further increases in $E/E_D$ and saturates at a value of about 0.3. We can see from Eqs. \ref{E_parallel}  and \ref{dreicer} that for a given temperature,
\begin{equation}
\frac{E}{E_D} \propto \frac{v_0 B_0}{n}.
\label{E_ratio}
\end{equation}
The local temperature at the acceleration site $n$ will depend on the details of the reconnection process and its determination would require a full analysis of the local dynamics. It is a reasonable assumption however that it scales with the density of the plasma flowing into the reconnection site such that $n \propto N$. We determine the constant of proportionality empirically by fitting the predicted radio power to that observed for the solar system planets, and find it to be of order 7. Hence, as shown in Fig \ref{E_ED}, the ratio $E/E_D$ is the same for all the solar system planets since the solar wind velocity is a constant and both the density and field strength vary as $R^{-2}_{\rm orb}$. This is not the case for close-in exoplanets, where the velocity, density and field strength vary differently with orbital radius. 

Electrons that are accelerated over a lengthscale $R_m$ to an energy $K = qER_m$ carry a power
\begin{equation}
P = \pi R_m^2 v N_{run} K [\mbox{W}] 
\end{equation}
where the velocity $v$ is either the wind velocity (for the outer planets) or the relative velocity of the star and planet (for the inner planets). Hence $P  \propto vN^2 R_m^2$ and so the dominant factors determining the output power are the size of the planetary magnetosphere and the density.  Using expressions \ref{density_m} to \ref{Rm_outer}, we can show that for given planetary parameters, this power scales as
\begin{equation}
P  \propto N_c^2 vB_\star^{-2/3}
\end{equation}
for the inner planets and
\begin{equation}
P  \propto N_w^2 vR_\star^4 R_{orb}^{-10/3}
\end{equation}
for the outer planets. Both of these cases are shown in Fig. \ref{power_all}. For those planets embedded in the stellar wind, the fall-off in power with increasing orbital radius is very pronounced. For the inner planets, however, the power is almost independent of orbital radius. As the orbital radius decreases, the number of electrons available to be accelerated rises (since $N_{run} \propto n^2 \propto R_{orb}^{-2}$). This is balanced however by the shrinkage of the collecting region of the magnetosphere that is apparent in Fig. \ref{size} since $R_m^2 \propto R_{orb}^{2}$. 

The power carried by the accelerated electrons for these inner planets is therefore largely independent of the planetary orbital radius (albeit with a weak variation due to the relative velocity $v$). The magnitude of this power is however affected by the stellar parameters we have chosen. For a stellar field strength of only 1G, not only is the typical electron energy of the runaway electrons reduced to a few keV, (and so we set the characteristic electron energy to 1eV) but also since the electric field generated is in many cases less than the Dreicer field, fewer electrons are accelerated. A higher stellar field strength would of course provide a larger electric field and hence accelerate electrons to higher energies, but it would lead to a greater compression of the planetary magnetosphere and hence reduce the upper limit for the lengthscale over which these electrons could be accelerated. 

We can quantify the effect of the magnetic field by determining the minimum stellar field strength needed to ensure that the distance $L_{\rm min}$ required for electrons to be accelerated above (say) 10keV is less than the maximum available lengthscale (R$_m$). Thus
\begin{equation}
B_{\star,min}  \propto \frac{R_{orb}^{3/2}}{R_\star^3}
\end{equation}
for close-in planets, while for the outer planets
\begin{equation}
B_{\star,min}  \propto \frac{R_{orb}^{5/3}}{R_\star^3}.
\end{equation}
We show this minimum stellar field strength in Fig. (\ref{B_min}) where we assume a stellar orbital period equal to that of HD189733 for close-in planets and a stellar wind speed of 500kms$^{-1}$ for the outer planets. Clearly, only a small increase in the stellar field strength above 1G  would be adequate to ensure that a pool of electrons of energies greater than 10keV would be available to power the electron-cyclotron maser instability for most exoplanetary systems. Even below this limit, energies of greater than 1keV are easily achievable.

\section{Comparison with the solar system}
\begin{figure}
  \begin{center}
       \psfig{figure=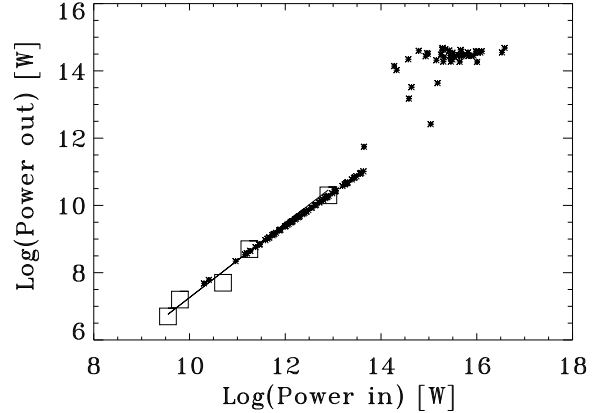,width=3.5in}
   \caption{Radio power assuming the electron cyclotron maser instability converts power in accelerated electrons into radio power at $10\%$ efficiency. Values for the solar system planets (shown as open squares) are taken from \cite{zarka_07}. A best-fit line through these values is shown. A stellar field $B_\star =1$G is assumed.}
   \label{Pout_Pin}
  \end{center}
\end{figure}

While Fig. \ref{power_all} illustrates the scaling of the expected power with orbital radius, the level of this power depends on our assumptions for the stellar field strength and densities. The fraction of this power that may be converted into observable radio emission also depends on the details of the electron distribution and the electron cyclotron maser instability (see e.g. review by \citep{wu_review_ecmi_85}). Values of $1 - 20 \%$ have been claimed \citep{louarn_92,mackinnon_ECM_92,cairns_CMI_02,vorgul_CMI_05} and so we assume that $10\%$ of the power in accelerated electrons is converted into radio power. This allows us to compare our predicted radio power with that observed from the planets in our own solar system,
which is directly proportional to the input magnetic power carried by the solar wind (see Fig. (\ref{Pout_Pin})). This ``radio-magnetic Bode's law'' has a  constant of proportionality or efficiency of $1-10 \times 10^{-3}$ relating the magnetic and radio powers \citep{zarka_07}. We can write this efficiency $\epsilon = P_{\rm radio}/P_{\rm magnetic}$ as
\begin{equation}
\epsilon = 0.1 \frac{\mu N_{\rm run} K}{B^2}
\end{equation}
where $K$ is the typical energy acquired by the accelerated electrons. The factor of 10$\%$ is an estimate of the fraction of the power in accelerated electrons that can be converted to radio emission.  Since we expect that $N_{\rm run} \propto N^2$, this gives
\begin{equation}
\epsilon \propto  \left[\frac{N}{B}\right]^2.
\end{equation}
For the outer planets (such as those in our own solar system) we can easily reproduce this constant efficiency $\epsilon$ by noting that both $N$ and $B$ scale as $(R_{orb}/R_\star)^{-2}$ This gives a constant efficiency whose magnitude depends principally on the local field strength and density. If we scale these to their values at the Earth's orbit (a density of $1.7\times10^{-20}$kgm$^{-3}$  and a field strength of 3.5nT) then we can reproduce the emission from the solar system planets shown in Fig. \ref{Pout_Pin}. 

%---------------------Fig 9------------------------
%------------------------------------------------------------

\section{Discussion}

For the exoplanets that are far from their parent stars, a universal ``radio-magnetic Bode's law'' might therefore be expected, but with a slope that is determined by the ratio of $N/B$ in the stellar wind. These two factors may well not be independent and indeed may vary with both spectral type and stellar age. \citet{griessmeier_planet_winds_05} have considered the effect on the detectability of radio emission of changes in stellar winds as stars age. They conclude that young systems with more powerful winds are the most likely candidates. 

The inner planets are, however, much more promising candidates for detection because of their larger predicted radio powers. The magnitude of this power also depends on their coronal density and to a lesser extent the stellar field strength. Increasing B$_\star$ increases the input power and so moves all the points in Fig. \ref{Pout_Pin} to the right. The output power, however, is not significantly affected. Thus, for example, if we consider HD189733 we can convert the radio power to a flux
%---------------------Fig 10------------------------
 \begin{figure}
  \begin{center}
       \psfig{figure=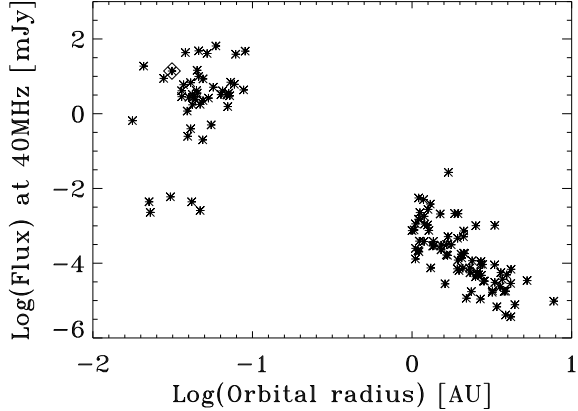,width=3.5in}
   \caption{Planetary flux at 40 MHz (in mJy) received at Earth. A stellar field $B_\star =1$G is assumed. In order of increasing orbital radius, the systems that lie above 10mJy are: Gliese 876 d, HD 189733 b, 55 Cnc e, HD 179949 b, Tau Boo b, 51 Peg b, Ups And b, HD 69830 b, and HD 160691 d. }
   \label{Flux_planet}
  \end{center}
\end{figure}
%------------------------------------------------------------
$\Phi$ received at Earth following \citet{farrell_radio_planets_99} and \citet{griessmeier_planet_winds_05, griessmeier_winds_07} to obtain
\begin{equation}
 \Phi = \frac{P}{\Omega d^2 \delta f}.
\end{equation}
Here d is the distance to the star and we assume planetary properties similar to Jupiter \citep{zarka_conf_04}. Hence  $\delta f$ is the bandwidth, taken as $\delta f = f_c$ where $f_c  = 40$MHz is the maximum cyclotron frequency and we assume that the emission is beamed into a solid angle $\Omega$ = 1.6 sr. While for a stellar field of 1G (as shown in Fig. \ref{Flux_planet}), the flux from HD189733 is only 14mJy, if we use the observed stellar field strength of 40G, the electric field increases so that the typical electron energy rises, but the magnetosphere shrinks, with the result that the flux changes to only 15mJy. This assumes, however, that the coronal density is equal to that of the Sun. If we assume scalings in the range $N \propto \Omega^{0.6}$ to  $N \propto \Omega$ \citep{unruh97loops,ivanova_taam_braking_03} then for the rotation rate of 11.73 days appropriate for HD189733, the flux rises to from 15mJy to 39 - 72 mJy respectively.

This example points to the importance of the stellar coronal density in determining the possible radio power. It suggest that those stars with high X-ray flux (which like the radio flux also scale as $N^2$) are the most likely candidates. Although rapid rotation may therefore lead to a greater radio power, most radial-velocity searches for exoplanets have tended to exclude rapidly-rotating objects from their survey samples, because their high rotational broadening degrades the precision with with radial velocities can be determined, and their surface activity causes radial-velocity ``jitter''.  Stars that are magnetically active tend to have significant radio emission of their own, which could mask any planetary signature \citep{benz_guedel_94}. Indeed, we might expect planet-induced emission due to the electron-cyclotron maser process at the star. This would appear at the cyclotron frequency of the stellar magnetic field $f [MHz]= 2.8B_\star[G] $ \citep{dulk85}. Thus, for HD189733 with an observed field strength of 40G \citep{moutou_HD189733_07} we might expect emission at 112MHz. In order for this emission to escape from the stellar magnetosphere, however, the local cyclotron frequency must be greater than the plasma frequency \citep{dulk85,bingham_horseshoe_00,zarka_07,griessmeier_07}. This means that at the site of emission,
\begin{equation}
B >  3 \times 10^{-6} N_e[m^{-3}] ^{1/2}  [\mbox{G}].
\end{equation}
For emission from the planet, this condition is easily met, requiring only a planetary field strength of 0.1G for a density of 10$^9$m$^{-3}$. If we are considering emission from the star, however, this condition requires that $B_\star > 60$G for a density of $4\times 10^{14}$m$^{-3}$. 

Stars that have higher field strengths will have a higher emission frequency and so may be easier to observe. In principle, this could be distinguished from the intrinsic emission of the star \citep{kellet_magnetic_trap_02} by its modulation due to the orbit of the planet, but this would require observation over many orbits to average out the effect of transient stellar emission due to stellar coronal activity. This process may also lead to chromospheric signatures of the interaction as electrons from the reconnection site are accelerated along the stellar field lines and impact the stellar surface. \citet{shkolnik03,shkolnik_05,shkolnik_spi_08} have reported planet-induced chromospheric activity on HD 179949 and $\upsilon$ And. \citet{mcivor_planet_06} modelled this process and showed that this interaction could explain the chromospheric activity, but not the phase lag between the planet and the chromospheric hotspot for $\upsilon$ And.  \citet{preusse_05,preusse_06} have, however, reproduced this phase lag by considering the propagation of Alfv\'en waves generated by the passage of an exoplanet through a stellar magnetic field.

The level of magnetic activity may be important in determining the rate at which the stellar field changes may also be important. These changes are most likely to be driven by new flux emergence through the stellar surface, or by surface transport processes: diffusion, differential rotation and meridional flows.  While there is now a general acceptance that surface differential rotation decreases with decreasing mass \citep{barnes_diffrot_05}, the nature of the surface diffusion or meridional flows is as yet uncertain. 

The stellar magnetic field also has a direct effect on the emission process through its control of the size of the planetary magnetosphere. Indeed, if the stellar field strength is too large, the planetary magnetosphere may be totally crushed. We can calculate the minimum stellar field strength required to do this by setting the size of the planetary magnetosphere equal to the planetary radius in the condition for pressure balance (\ref{press_bal}), and setting the ram pressure term to zero. This gives
\begin{equation}
B_{\star,\rm min}  = 3.8 \times 10^3 \left(\frac{R_{orb}}{0.03 AU}\right)^3 \left(\frac{R_\odot}{R_\star}\right)^3  [\mbox{G}].
\end{equation}

For the subset of planets that transit their host stars, there is the possibility of observing radio occultations as the planet passes behind the star. With the observation of repeated occultations any time-dependence in the emission due to variations in the emission process itself could be averaged out. Such observations might provide information about the structure and orientation of the planetary magnetic field, since most of the emission is expected to originate in the region of the magnetic pole close to the planetary surface. Even if this emission is present, however, it may be beamed out of the line of sight.  Recent claims of electron-cyclotron emission from low-mass stars with kG magnetic fields are, however, extremely encouraging and suggest that occultations are indeed possible as the source region rotates out of view behind the star \citep{hallinan_08,antonova_hallinan_08}. The emission is at much higher frequencies than would be expected from a  planetary magnetic field however.

%There are clearly many parameters that can affect the likelihood that radio emission from an exoplanetary system will be detectable. Some of these can be determined observationally, but others can only be estimated, using our own solar system as a guide.

%\subsection{Planetary properties}
%The principle unknowns are the magnetic moment and rotation rate of the planet and also the density and temperature of the planetary exosphere which may themselves be influenced by the presence of satellites such as Io which can contribute to the emission process \citep{winglee_planet_radio_86,farrell_radio_planets_99}. Modelling of planetary exospheres and their interaction with a stellar wind is ongoing \citep{griessmeier_planet_tides_04,lipatov_planet_winds_05,jaritz_planet_roche_05,griessmeier_planet_winds_05,stevens_planet_winds_05} but observational confirmation of the conditions within planetary exospheres can probably only be obtained for transiting systems.

\section{Conclusions}
 In studying the possible radio emission from exoplanets, we have focussed on the electric field that is generated when the planetary and stellar magnetic fields reconnect. This electric field can accelerate electrons to energies sufficient to power the electron-cyclotron maser instability which is believed to be responsible for much of the radio emission from the planets in our solar system. We have assumed that the mass, radius and magnetic field strength of all exoplanets is the same as Jupiter, but used the observed stellar values. By modeling the electric field developed in this interaction, we have calculated the number of accelerated electrons produced. We find that it scales as the square of the density $N$. Thus the output radio power scales as $N^2$ while the input power from the stellar magnetic field scales as $B^2$. Since for planets immersed in a stellar wind (as is the case for the planets in our own solar system) both the density and the field strength scale as $R_{\rm orb}^2$, this naturally predicts that the output and input power should be proportional, as is observed for the solar system. 

We find, however, that the output power from the inner planets behaves differently. The planets lie inside the closed magnetic field (or magnetosphere) of the star. The magnetospheres of these planets are confined not by the ram pressure of the stellar wind, but by the magnetic pressure of the stellar magnetosphere. The density and magnetic pressure within which they are immersed do not vary with orbital radius in the same way as for the outer planets, and so we would not expect the output and input powers to be proportional. In fact, we find that as the planetary orbital radius decreases, the density rises and so the pool of available electrons increases, but this is exactly balanced by the shrinking of the magnetosphere which reduces the cross-section of the planet. The result is that while the input power might rise, the output power saturates at a value determined by the ratio 
\begin{equation}
P_{\rm out} \propto \left[ \frac{N_c}{B_\star^{1/3}}\right]^2
\end{equation}
where $N_c$ is the density at the base of the stellar corona and $B_\star$ is the surface field strength. We conclude, therefore, that the radio power emitted by ``hot Jupiters" is determined by the way that coronal density scales with stellar field strength as a function of stellar rotation rate and effective temperature.

 %----------------------------------------------------------------------
%\section{Acknowledgements}

%The authors would like to thank Aad van Ballegooijen who wrote the original version of the field extrapolation software. 

%--------------------------------------------------------------------

%\bibliographystyle{aa}
%%\bibliography{iau_journals,master,ownrefs,mmjpapers}
%\bibliography{iau_journals,mmjpapers}
%\bibliography{mmjpapers}

%----------------------------

%-----------------------------
\appendix
\section{Modelling the reconnection of the stellar and planetary magnetic fields}

\begin{figure}
\centering
    \psfig{figure=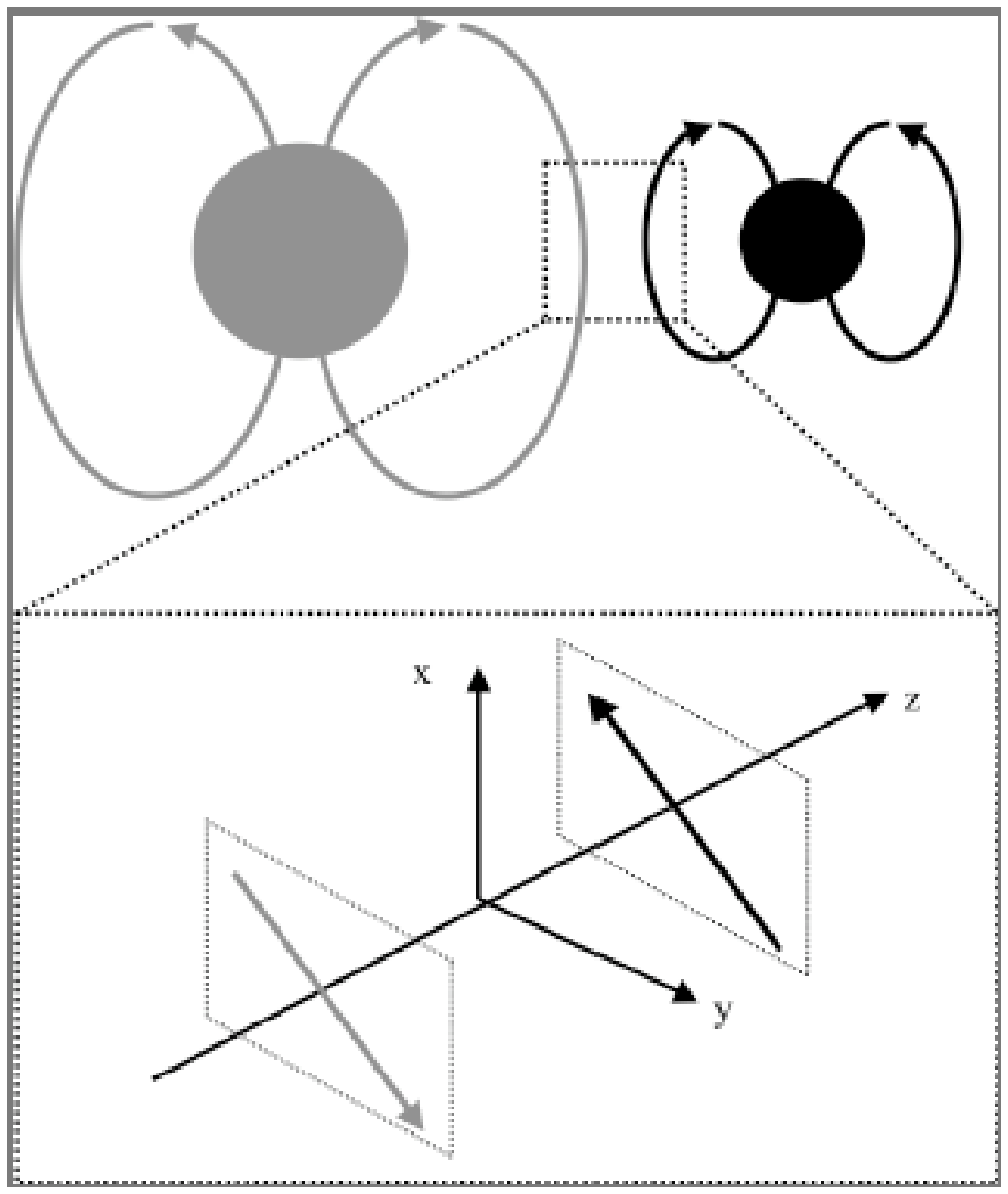,width=3.5in}
\caption{Schematic diagram of the coordinate system used for the magnetic field
at the centre of the current sheet where the stellar (grey) and planetary (black) magnetic fields reconnect. The
z-axis points along the line of centers joining the star and planet.}
\label{coords}
\end{figure}

We model the interaction of the stellar and planetary magnetic fields
in the manner similar to that used for colliding wind systems \citep{jardine96particle}. Because
of the large length scales involved we neglect the radius of curvature
of the reconnecting field lines. In this case, the equation of motion
of the flow is just
\begin{equation}
\label{eqn:of:motion}
  \rho({\bf v}\cdot{\bf \nabla}){\bf v}  =   - {\bf \nabla}
    \left[
      p +B^2/2\mu
    \right]
  + \rho\nu\nabla^2{\bf v},
\end{equation}
where $\nu$ is the kinematic viscosity and the other symbols have their
usual meaning. For this local analysis, we use a Cartesian coordinate system where the $z$-axis lies 
along the line of centres of the star-planet system (see Figure \ref{coords}). As the magnetic fields of the star and planet are pushed together by the local flow, a current sheet is generated. Within this current sheet, components of the magnetic field that are antiparallel will annihilate, while components that are parallel will combine. We therefore chose to align the $x$- and $y$-axes such that the
$x$-components of the magnetic field on either side of the current sheet
are exactly antiparallel and hence cancel out within the current sheet, 
while the $y$-components  are parallel. As a result, at the center of 
the current sheet the magnetic field has only a $y$-component. The 
magnetic field is then of the form
\begin{equation}
{\bf B} = (B_x(z),B_y(z),0)
\end{equation}
and the current density $\bf{j} = \bf{\nabla}\times\bf{B}/\mu$ is
\begin{equation} 
\mu{\bf j} = (-B'_y, B'_x, 0 ).
\end{equation}
The variation with distance from the centre of the current sheet of the velocity and magnetic field for colliding magnetofluids
has been determined for arbitrary vorticities by \citet{jardine93},with a numerical
solution that shows that the magnetic field strength at the centre
of the current sheet varies only slowly with vorticity. We therefore consider only
the case of a zero-vorticity flow for which an exact solution for the 
magnetic field profile can be found  \citep{sonnerup75}.
In this case, the velocity is of the form
\begin{equation}
{\bf \overline{v}} =  (\overline{x},\overline{y},-2\overline{z})
\end{equation}
where all distances are scaled to some global lengthscale $L$ so that $\overline{x}=x/L$ etc and
$\overline{v}=v/v_0$ where $v_0$ is the velocity with which field lines are
being carried towards the current sheet. Now, by our assumption of a steady 
state, $\bf{\nabla}\times \bf{E} = 0$ where the electric field is given by a simple Ohm's law
\begin{equation}
\bf{E} = -\bf{v}\times\bf{B} + \bf{j}/\sigma.
\end{equation}
This can be written as  
\begin{equation}
 \label{B}
        \frac{\overline{B}''}{H}
      + 2\overline{z}\overline{B}'
      + \overline{B} =0,
\end{equation}
where $H=v_0 L/\eta$ is the magnetic Reynolds number and 
$\overline{B}=B/B_0$ where $B_0 = B(\overline{z} = 1)$.
The electric field $\overline{E} = E/v_0 B_0$ has components
\begin{equation} 
       {\bf \overline{E}} = (-2 \overline{z} \overline{B}_y - 
\overline{B}'_y/H,
                               2 \overline{z} \overline{B}_x + 
\overline{B}'_x/H,
                      - \overline{x} \overline{B}_y + \overline{y} 
\overline{B}_x). 
\end{equation}
If we take the component of $\bf{E}$ along $\bf{B}$ to be 
$\bf{E}_{||}$, it is straightforward to show that
\begin{equation}
{\bf E}_{||} = {\bf E}\cdot {\bf B} 
                        = {E}_{||}(0) e^{-z^2H}
\end{equation}
and so we would expect that the centre of the current sheet is the 
site of the most efficient particle acceleration. We therefore
concentrate our attention on the particles accelerated there. 
 The 
degree to which the magnetic field rotates as it comes into the
current sheet then depends on the relative magnitudes of the 
x- and y-components outside the current sheet. We specify
this as one of the boundary conditions, in fact making them equal
at $z=R_m$ so that the field rotates through $\pi/4$ radians. 
With boundary conditions 
\begin{equation}
  \label{mag_bcs}
  \overline{B}_{\rm{x}}(0)=0, \makebox[0.25cm]{}
  \overline{B}_{\rm{x}}(1)=1, \makebox[0.25cm]{}
  \overline{B}'_{\rm{y}}(0)=0, \makebox[0.25cm]{}
  \overline{B}_{\rm{y}}(1)=1
\end{equation}
the solution to (\ref{B}) can be expressed in terms of modified Bessel
functions as
\begin{equation}
    \overline{B}_{\rm{x}} = \frac{z^{1/2}e^{(1-z^2)H/2}}{I_{1/4}(H/2)}
                  I_{1/4}(z^2 H/2)
\end{equation}
and
\begin{equation}
    \overline{B}_{\rm{y}} = \frac{z^{1/2}e^{(1-z^2)H/2}}{I_{-1/4}(H/2)}
                  I_{-1/4}(z^2 H/2)
\end{equation}
For $H>>1$ we can write
$I_{\pm1/4}(H/2) \approx \left( e^{H}/ \pi H \right)^{1/2}$
so that at the centre of the current sheet we have
\begin{equation}
    \overline{B}_{\rm{y}}(0) = \frac{\sqrt{2\pi}}
                                       {\Gamma(3/4)} H^{1/4}
\label{odd}
\end{equation}
\begin{equation}
   \overline{B}'_{\rm{x}}(0) = \frac{ 2\sqrt{2\pi}}
                                      {\Gamma(1/4)}H^{3/4}
\end{equation}
and
\begin{equation}
     \overline{E}(0) = -\frac{ 2\sqrt{2\pi}}
                             {\Gamma(1/4)} H^{-1/4}
\end{equation}
or
\begin{equation}
    {E}(0) = -\frac{ 2\sqrt{2\pi}}
                             {\Gamma(1/4)} H^{-1/4} v_0 B_0.
\end{equation}
%-----------------------------------------------------------------------------
%------------------------------------------------------------------------------

\end{document}